\documentclass[fleqn,twoside]{article}
\usepackage{espcrc2}

\usepackage{graphicx}
\usepackage[figuresright]{rotating}

\newcommand{\AmS}{{\protect\the\textfont2
  A\kern-.1667em\lower.5ex\hbox{M}\kern-.125emS}}

\title{Relic Densities of Gauged Axions and Supersymmetry}

\author{Claudio Corian\`o\address[MCSD]{Dipartimento di Fisica\\ Universit\`{a} del Salento and INFN, Via Arnesano, 73100, Lecce, Italy }%
        \thanks{ Presented at NOW 2010, Conca Specchiulla, 4-11 September 2010, Otranto (Lecce), Italy. },
        Marco Guzzi\address{Department of Physics, Southern Methodist University, Dallas TX 75275, USA},
         Antonio Mariano\addressmark[MCSD]
       }
       
\begin{document}

\begin{abstract}
We illustrate the structure and the main phenomenological features of a supersymmetric model (the USSM-A) built following a bottom-up approach and containing an anomalous abelian gauge symmetry. This model supports a gauged axion in its spectrum and provides a generalization of the global (supersymmetric) Peccei-Quinn construction. Complete simulations of the neutralino relic density are performed. Bounds from CAST and WMAP, combined with dark matter simulations, provide significant constraints on the scale of the interactions between the axion and the gauge fields. 

\vspace{1pc}
\end{abstract}

% typeset front matter (including abstract)
\maketitle

\section{Introduction}
Gauged shift symmetries are typical of potentials containing flat directions and are common to several theories built around the Planck scale, such as strings/branes, down to supergravity and supersymmetric theories. The latter inherit the rich structure of brane models via some mechanism of dimensional reduction or geometric compactification, in the presence of external fluxes.  However, the dynamics of scalar fields (massive and massless), associated to these flat directions, which take the form of massive moduli and massless Goldstone modes, is quite involved. This is in part due to the excessive proliferation of scalars in the low energy theory. 

On the other hand, one could also take into consideration the possibility that some of these flat directions could be (almost) preserved as the Universe expands, down to the low energy scale. In particular, they could be slightly lifted (non-perturbatively) from their unperturbed vacuum value at the electroweak and QCD phase transitions. In this case one could envision possible contributions to the dark matter density from these very light scalars/pseudoscalars. 

%At these scales, a small curvature could be induced, thereby giving a small mass to these Goldstone modes.

Although rather exotic at a first glance, this picture is not new, since it has been at the core of the long known Peccei-Quinn (PQ) proposal for the solution of the strong CP-problem, that shares some of its typical features.  

According to this mechanism, which invokes a very light pseudoscalar in the physical spectrum, a global anomalous $U(1)$ symmetry is attached to the fields of the Standard Model, which breaks at a very large scale 
($f_a\sim 10^{12}$ GeV). In turn, the nature of the axion, as a (pseudo) Nambu-Goldstone mode of the broken global $U(1)$, has to be found at a later stage in the Early Universe, at the QCD transition ($\sim \Lambda_{QCD}$), with the generation of a periodic potential due to the instanton vacuum.

The rather singular nature of this mechanism, which involves two widely separated scales, is made evident by the expression of the axion mass ($m_a$) which is related to their ratio $(m_a\sim\Lambda_{QCD}^2/f_a)$ and by its rather small value ($m_a\sim 10^{-3}$ eV). 

In a supersymmetric context, the axion is the imaginary component of a complex scalar ($b$), and is accompanied by another degree of freedom, the saxion, described by $\textrm{Re}\, b$,  and by a supersymmetric fermionic partner (the axino, $\psi_b$). The gauging of this multiplet has been discussed in \cite{Kors:2004ri}, and further in \cite{Feldman:2010wy} in a study of the St\"uckelberg extension of the MSSM, with a single extra (non anomalous) $U(1)$ symmetry. In this extension $\textrm{Im}\, b$ is a Goldstone mode of the $U(1)$ gauge symmetry and, as such, becomes the longitudinal component of the $U(1)$ gauge field, disappearing completely from the physical spectrum.  A similar destiny is shared by the axion of a $U(1)'$ (prime) extension of the MSSM (U(1)' MSSM) \cite{Anastasopoulos:2008jt}, which is anomalous and introduces extra interactions in the form of PQ counterterms for the restoration of gauge invariance (supersymmetric terms which generalize the $\textrm{Im}\, b\, F\tilde{F}$ vertex). In this second model the axino mixes with the gauginos and Higgsinos of the theory to generate several neutralinos and the corresponding LSP (light supersymmetric particle).

The first model that supports a physical axion and is compatible with supersymmetry, called the USSM-A, is built around the structure of the USSM \cite{Coriano:2008aw,Coriano:2008xa}, with some important variants: 1) there is no extra Higgs to ensure the breaking of the 
$U(1)$ symmetry, rather, this is realized in the St\"uckelberg form; 2) the $U(1)$ symmetry is anomalous. These two conditions, together with a requirement on the charge assignments of the Higgs sector that mixes the St\"uckelberg and the Higgs mechanisms, allow to 
construct a complete supersymmetric model where the gauged axion is a component of $\textrm{Im}\, b$. 

\subsection{The physical axion}
The extraction of a physical axion in theories of this type was pointed out in \cite{Coriano:2005js} in a non-supersymmetric context, motivated in a certain class of string vacua, and in successive phenomenological studies \cite{Coriano:2007fw,Coriano:2007xg}.
 It was later remarked that these effective actions could be generated starting from anomaly-free theories, with no reference to any class of string vacua, under particular conditions on the decoupling of a chiral fermion.  An example of such a behaviour is encountered when a heavy Higgs decouples from the low energy spectrum, "dragging" away also a part of the fermion spectrum (via Yukawa interactions) and one gauge boson, which becomes very massive \cite{Coriano:2009zh}.

 The effective low energy theory carries the signature of this "partial decoupling" of the Higgs, inheriting effective interactions which are PQ like. In this case, as in the original PQ model, the low energy axion is related to the phase of the heavy Higgs field.
 
Before coming to the issue of the mass of the axion in this model, we should mention that this is generated by an extra potential, allowed by the gauge symmetry \cite{Coriano:2005js}, which is periodic in the physical axion field \cite{Coriano:2010ws,Coriano:2010py}. The potential is gauge invariant only if the the extra singlet superfield $\hat{S}$ of the superpotential is charged under the anomalous $U(1)$. In turns, this implies that the two Higgs superfields of the model are also charged under the same $U(1)$ and are not charge-aligned. As a result of this choice, the mass of the anomalous gauge boson is induced both by the Higgs and the St\"uckelberg mechanisms.  We briefly comment on other essential features.
 
1) The anomalous fermion spectrum induces trilinear gauge interactions which are absent in anomaly-free extensions. These consist of cubic $(U(1)_B^3)$ anomalies and mixed-anomalies of $U(1)_B$ with the hypercharge gauge field (Y), and in combination with non-abelian anomalies with the $SU(2)$ and $SU(3)$ gauge fields. 2) For a general anomalous interaction, the D-terms of the scalar potential 
are non-local, but expandable in the St\"uckelberg scale. This feature can be avoided by a suitable choice of the fermion charges, without causing a complete decoupling of the extra $U(1)_B$. 3) The (St\"uckelberg) mass of the anomalous gauge boson, $M_{St}$, is a free parameter of the theory 
and is the suppression scale of the $\textrm{ Im}\, b\, F \tilde F$ interactions.

\section{Phenomenology}
In the supersymmetric case $M_{St}$ has necessarily to lay around the susy breaking scale in order to comply with the WMAP bounds \cite{Hinshaw:2008kr} on the neutralino relic densities (see Fig. 1). 
In fact, if it is taken to be too large, a mechanism of see-saw in the neutralino sector appears to be inevitable in this model \cite{Coriano:2010ws}, causing the LSP (neutralino) to be too light. In this case, the thermal decoupling of the neutralino is  not accompanied by a significant co-annihilation that could lower the corresponding relic densities, and this would be sufficient to 
violate the WMAP constraints on the allowed contribution to the relic density $\Omega h^2$ coming from dark matter ($\sim 10-14 \%$). This bound is present only in the supersymmetric scenario \cite{Coriano:2010py}.

Coming to the mass of the axion in this model, this depends on the size of the extra potential $V'$, generated at the electroweak scale \cite{Coriano:2010py}. This remains, in our
construction, undetermined.
 In particular, if the
extra potential is generated by non-perturbative effects at the
electroweak phase transition, then its mass is tiny, and the true mechanism of misalignment which determines the mass value
takes place at a second stage, at the QCD phase transition. Notice that due to the presence of both SU(2) and SU(3) anomalies, 
we should allow sequential vacuum misalignments of this field. In this
case the axion would be quite similar to an ordinary PQ axion, but with an interaction with the photons which would be difficult to reconcile with the CAST experiment (in the supersymmetric case) due to the bounds on $M_{St}$ \cite{Arik:2008mq}.   

 A second possibility that we have investigated is based on the assumption that
the size of the extra potential is unrelated to
electroweak instantons. In this case the mass
of the axion remains a free parameter. The range that we have explored involves an axion mass in the MeV region. We have discussed in \cite{Coriano:2010ws}
the several constraints that emerge from the model. Then the
axion is, in general, not long-lived and as such is not a component of
dark matter. On the other hand, the constraints from CAST can be
avoided, since the particle would not be produced at the center of the sun.

\begin{figure}[t]
\vspace{9pt}
%\framebox[55mm]{\rule[-21mm]{0mm}{43mm}}
\includegraphics[scale=.65]{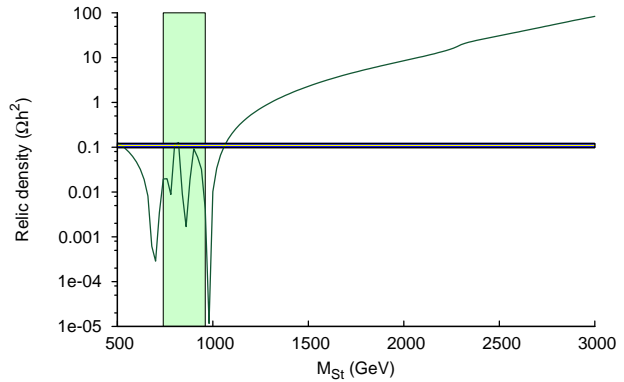}
\caption{Neutralino relic density ($\Omega h^2$) as a function of the St\"uckelberg mass $M_{St}$ in the USSM-A. The co-annihilation region is indicated by the shadowed area. The horizontal line is the WMAP constraint. }
\label{fig:largenenough}
\end{figure}
\centerline{\bf Acknowledgements} 
{We thank M. Masip for discussions and hospitality at the University of Granada. This work is partially supported by MICINN-INFN funds with the University of Granada}

\end{document}